%% file: revised-Incera-CSQCDIII-2013.tex
\documentclass[12pt]{article}
\usepackage{epsfig}
\def\beq{\begin{equation}}
\def\eeq{\end{equation}}

\def\bea{\arraycolsep .1em \begin{eqnarray}}
\def\eea{\end{eqnarray}}

\input econfmacros-csqcd3.tex
\def\Title#1{\begin{center} {\Large {\bf #1} } \end{center}}

\begin{document}

\Title{Magnetic Field Effects in Fermion Pairings}


\begin{raggedright}
{\it Vivian de la Incera\index{de la Incera, V.}\\
Department of Physics\\
University of Texas at El Paso\\
El Paso, TX 79968\\
USA\\
{\tt Email: vincera@utep.edu}}
\end{raggedright}
\section{Introduction}

The structure of neutron stars is directly connected to the equation of state (EoS) of their matter content. The strongly interacting matter at the core of a neutron star is mainly formed by protons and neutrons and is usually described by effective nuclear models. Nevertheless, for both strange stars that are formed by deconfined u,d and s quarks, or for hybrid stars with very dense cores, where the overlapping of the nucleons could lead to quark deconfinement, a description in terms of quark matter makes more sense. For these cases, QCD-like quark models have to be used to investigate the different phases that can be realized depending on the values of the density, temperature, and external fields present. Finding observable signatures that allow to distinguish among different internal phases of the neutron stars is a main goal in the astrophysics of compact objects.

The physics of neutron stars is hence intimately related to the investigation of the QCD phases, whose properties are explored in heavy-ion collisions at several experimental facilities all over the world. In this regard, an important problem to be investigated is the influence of a magnetic field on the structure of the QCD phase diagram, and particularly, on the location and the nature of deconfinement and chiral symmetry restoration. One reason for this interest is, on the astrophysical side, the existence of magnetic fields $\sim10^{14}-10^{15}$G, in the surface of magnetars \cite{Duncan}, with inner values estimated to be $10^{18}$ G or $10^{19}$ G, for hybrid stars with nuclear  \cite{ApJ383} or quark matter \cite{PRC82} cores respectively, or even higher, $\sim10^{20}$ G, for strange stars \cite{PRC82}. Close to home, the production of very strong magnetic fields in off-central heavy-ion collisions as in the Au-Au collisions at RHIC, can generate magnetic fields as large as $10^{18}$ G,  while fields even larger $\sim10^{19}$ G, can be generated with the energies reachable at LHC for Pb-Pb collisions. Even though these magnetic fields decay quickly, they only decay to a tenth of the original value for a time scale of order of the inverse of the saturation scale at RHIC \cite{Fukushima-2939}-\cite{PRD49}, hence they may influence the properties of the QCD phases probed by the experiment. Strong magnetic fields will likely be also generated in the future planned experiments at FAIR, NICA and JPARK, which will make possible to explore the region of higher densities under a magnetic field. 

Quite often, going from one QCD phase to another is connected to the change in the expectation value of a fermion condensate. For example, the order parameter for the chiral phase transition is a fermion-antifermion  condensate $\langle\overline{\Psi}\Psi\rangle$, while the order parameter for the transition from color superconductivity to normal quark matter is a fermion-fermion condensate of generic form $\langle\overline{\Psi}_C \Gamma\Psi\rangle$, with $\Gamma$ being some matrix in color, flavor, and Dirac space that depends on the model under consideration. At intermediate densities, a condensate of quarks and holes can be formed leading to an inhomogeneous phase on which chiral symmetry is broken in a different way. Each of these condensates comes from pairing interactions that are present in the original QCD theory and can be modeled in effective theories through Nambu-Jona-Lasinio-like interaction terms. The influence of an external magnetic field on these different condensates is relevant to understand how it will affect the QCD phases and the phase transitions.

In this paper, I shall discuss various fermion pairings that are of interest for the QCD phases, and explore the effects of an external magnetic field on these pairings, on the realization of new condensates, and on the properties of the magnetized phase. 

\section{Fermion Pairings in a Magnetic Field}
\subsection{Magnetic Catalysis of Chiral Symmetry Breaking}
\begin{figure}[!ht]
\begin{center}
\includegraphics[width=0.35\textwidth]{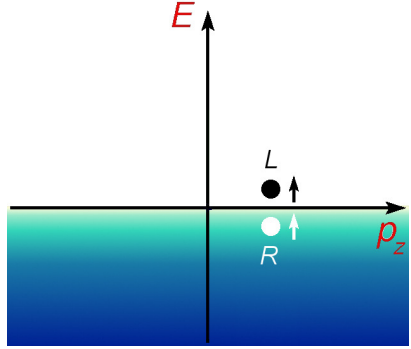}
\caption{Energy-momentum sketch of chiral pairing in vacuum.}
\label{chiralcondensate}
\end{center}
\end{figure}

The simplest pairing,  represented in Fig \ref{chiralcondensate}, takes place between particles and antiparticles at the Dirac sea. The particle and antiparticle have opposite spin, opposite chiralities, and opposite momenta. Hence, the chiral condensate is a neutral and homogenous scalar that breaks chiral symmetry. This is the well-known chiral condensate responsible for the constituent quark mass in QCD. It only occurs when the strength of the coupling between the fermions is stronger than some critical value which depends on the details of the interaction and the model considered. 

One may wonder how a magnetic field can influence a neutral condensate, but let's not forget that the fermions in the pair are charged, so each of them can minimally couple to the magnetic field. This coupling leads to the Landau quantization of the fermion's momentum. Recall that the energy of a fermion of mass $m$ and charge $q$ in a uniform magnetic field is given by
\beq \label{energyinB}
\sqrt{p^2_3+2qBn+m^2}, \quad  n=0,1,2,...
\end{equation}
where $B$ is the strength of the magnetic field and we have assumed, without lost of generality, that it points in the $z$ direction. The Landau level $n$ characterizes the quantization of the momentum in the direction transverse to the field. It is easy to see that the infrared dynamics of the particles in the lowest Landau Level  (LLL), $n=0$, is 1+1-dimensional. In the chiral limit there is no energy cost to excite the LLL particles about the Dirac sea and a large number of degenerate excitations are produced. This situation makes the system unstable against the formation of particle-antiparticle pairs, which are now favored even at the weakest attractive interaction. This is the well-known phenomenon of magnetic catalysis of chiral symmetry breaking (shorten as MC from now on) \cite{MC}. It is very similar to the  Bardeen-Cooper-Schiffer (BCS) mechanism that takes place about the Fermi surface and favors the formation of Cooper pairs. 

In the original works on MC, this mechanism was only associated with the generation of a chiral condensate that in turn leads to a dynamical fermion mass. However, it is easy to understand that in the presence of a magnetic field, a second condensate is unavoidable. To see this, notice that the fermion-antifermion pair possesses a net magnetic moment, because the particles in the pair have opposite charges and spins. In the absence of magnetic field, these magnetic moments point in all directions and have no effect on the ground state. But when a magnetic field is present, the pairs' magnetic moments orient themselves in the direction of the field, so the ground state can have a net component of the anomalous magnetic moment (AMM) in the field direction, that manifests as a spin-one condensate of Dirac structure $i\gamma_1\gamma_2$. This can also be seen as a consequence of the explicit breaking of the rotational group by the field. In the presence of the magnetic field only the subgroup $O(2)$ of rotations about the field axis remains as a symmetry of the theory. 

Given that the two symmetries the new condensate would break, chiral and rotational, are already broken, either spontaneously or explicitly, these symmetries are not protected, and nothing prevents the emergence of this spin-one condensate along with the conventional chiral condensate. The situation resembles the realization of the symmetric gaps in the Color-Flavor-Locked (CFL) phase of color superconductivity \cite{symmgap1}-\cite{MCFL}.  Depending on the theory considered, the presence of an AMM condensate manifests in the free energy as a term that couples the field with the dynamical AMM (QED) \cite{MCAMM}, or, for QCD-inspired NJL theories, through a spin-spin interaction generated via the Fierz identities in a magnetic field \cite{EV-inprep}. No matter how the existence of the AMM is manifested in the particular theory, one can always show that in the presence of a magnetic field, the ground state of the system does not admit a solution that has nonzero chiral condensate, but zero AMM.  Therefore, the existence of this spin-one condensate is unavoidable and universal in the MC phenomenon. An AMM condensate can produce effects like a nonperturbative Zeeman splitting in massless QED \cite{MCAMM} .

The MC is a very universal mechanism that has been corroborated in QED and in many different fermion model calculations in vaccum and at finite temperature. Nevertheless, some recent lattice QCD calculations in a magnetic field have produced contradictory results. While in Ref. \cite{latticeQCDMC} the validity of the MC behavior was corroborated, Ref.\cite{latticenoMC} claims that the MC scenario is found at low temperatures, but around the crossover temperature the chiral condensate shows a complex behavior with the magnetic field that results in a decrease of the transition temperature with the field. This interesting issue is still under scrutiny and more investigations will be needed to settle it. 

\subsection{Fermion Pairing at Finite Density in a Magnetic Field}
\begin{figure}[ht!]
\begin{center}
\includegraphics[width=0.35\textwidth]{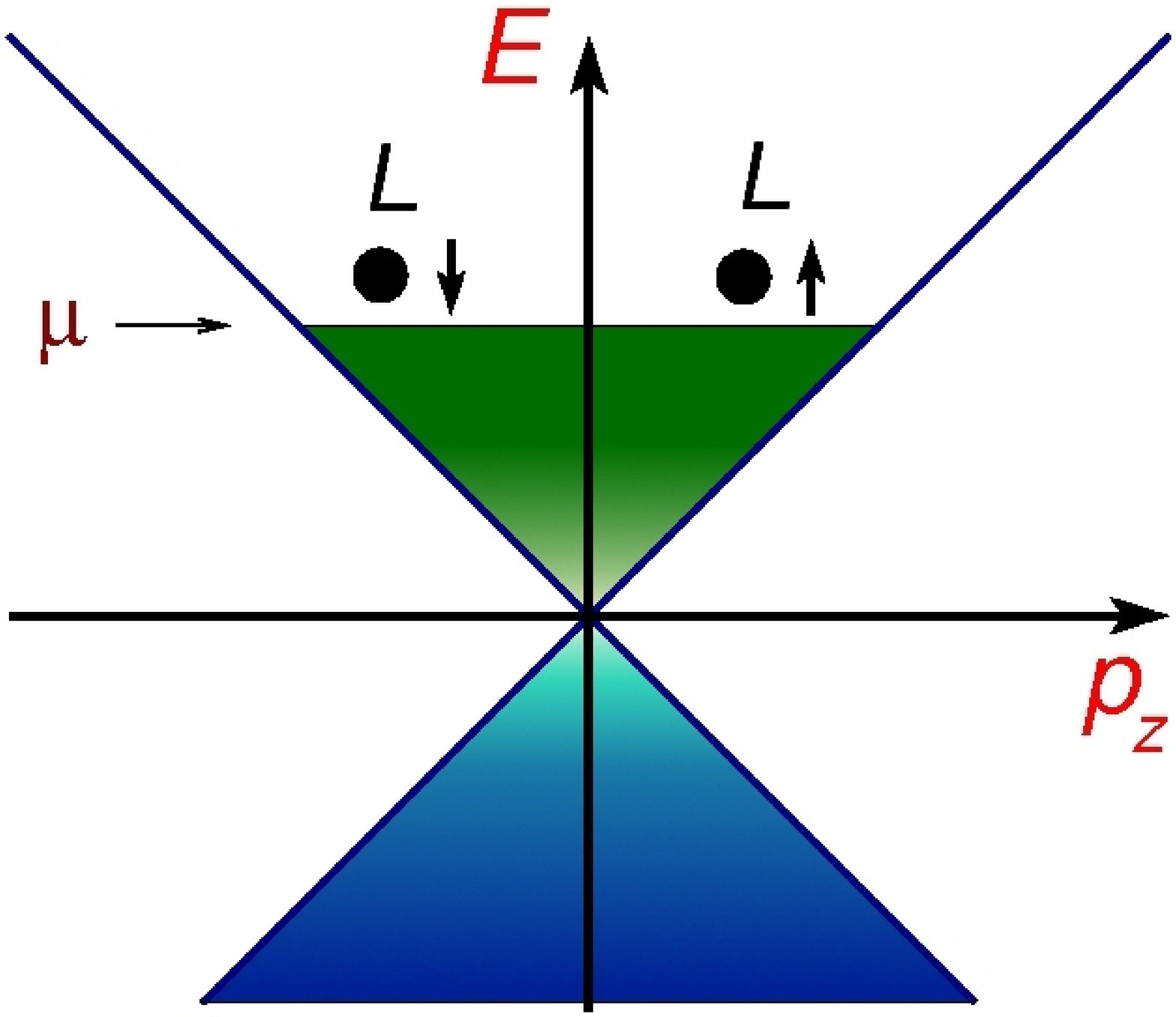}
\includegraphics[width=0.35\textwidth]{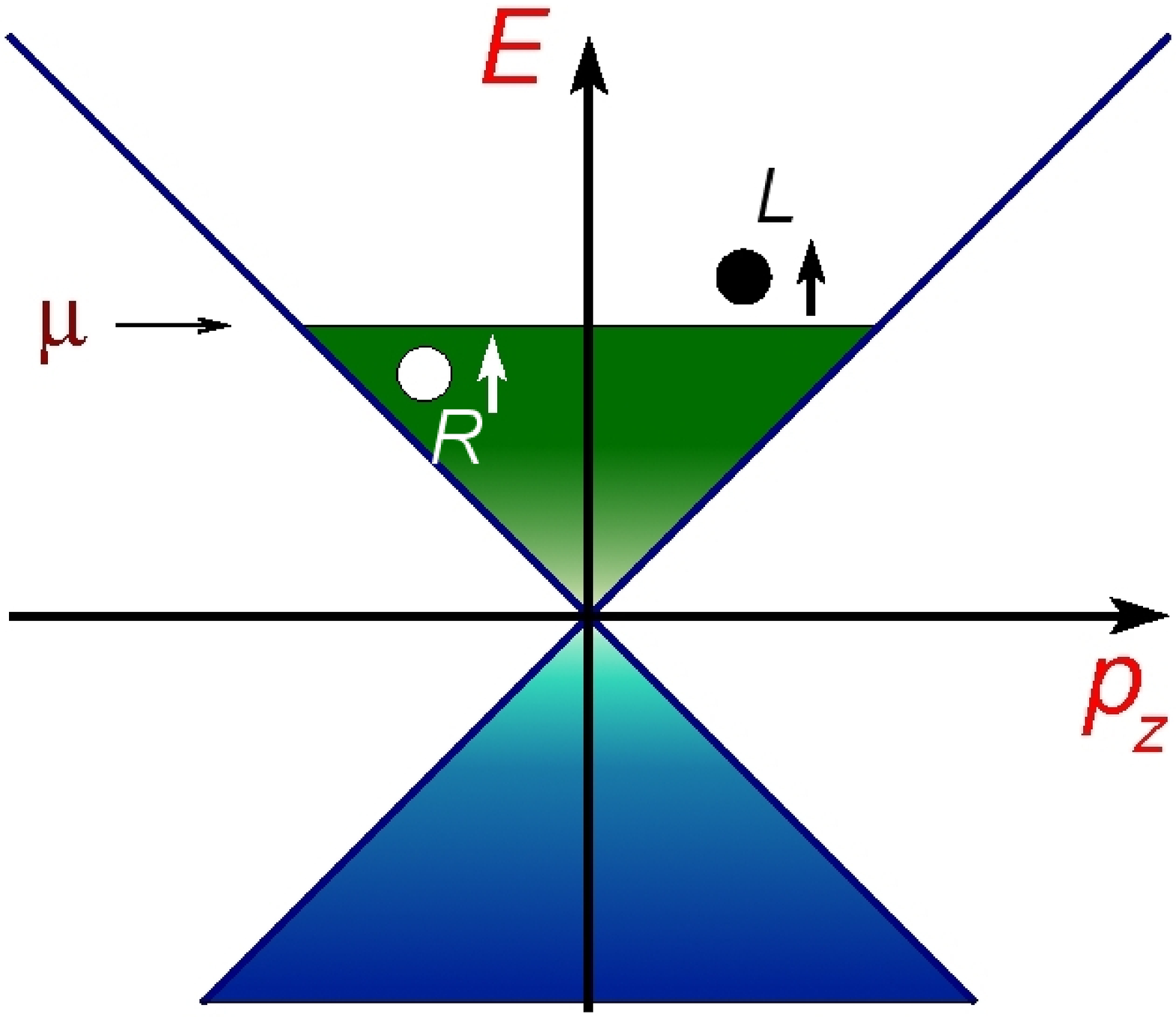}
\caption{\footnotesize Two possible pairings at finite density. Left panel: Cooper pairing that pairs fermions of the same chirality and opposite momenta and spins. It can only occur if a fermion-fermion attractive interaction, no matter how weak, exists. Left panel: Density wave pairing that pairs particles of opposite chirality. The pair is formed by a fermion of momentum P and a hole created by the absence of a fermion of momentum -P in the Fermi surface.} 
\label{pairingfinitedensity}
\end{center}
\end{figure}

At finite density the chiral condensate is less favored because hopping the antiparticles from the Dirac sea to the Fermi surface, where they can pair with the particles, costs twice the Fermi energy. In this case the magnetic field's influence on the chiral pair would only be important for fields much larger than the density. Two other pairings are however favored at finite density and can lead to very interesting new physics. One is the Cooper pairing responsible for the BCS superconductivity, occurring between fermions at the Fermi surface whenever an attractive interaction, no matter how weak, is present (left panel in Fig.\ref{pairingfinitedensity}). The other is the density wave (DW) type of pairing between a particle and a hole of momentum $\mathbf{P}$ each (right panel in Fig.\ref{pairingfinitedensity}). The DW pairing is familiar for two-dimensional systems in condensed matter. In four dimensions, the DW pairing requires a strong coupling to be favored over the BCS type. While the BCS condensate is homogenous, the DW one is inhomogeneous and hence breaks translational symmetry. These condensates are familiar in condensed matter, but they are also relevant in QCD at finite density. 

At very large densities the most favored pairing in QCD is of the BCS-type, because the effects of the quarks on the gluon screening become large, making the coupling weak and decreasing the likelihood of a DW condensate. On the other hand, the DW condensate may have an edge over color superconductivity in the region of intermediate densities, where not only is the coupling stronger, but also, since it pairs single-flavor quarks, is immune to the pairing stress produced by different quark chemical potentials that leads to chromomagnetic instabilities in color superconductivity.

\begin{figure}[ht!]
\begin{center}
\includegraphics[width=0.35\textwidth]{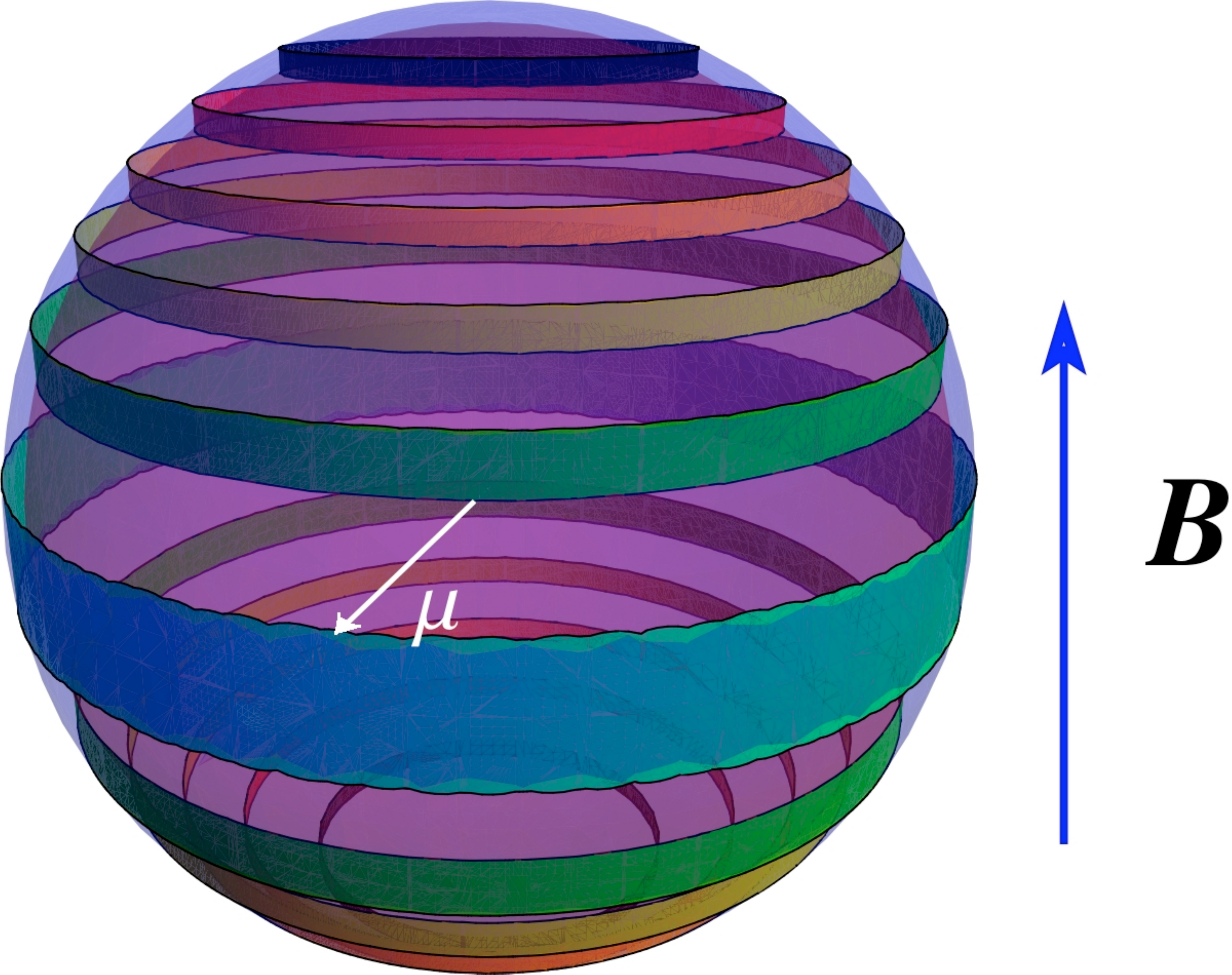}
\caption{\footnotesize Fermi surface in a magnetic field. } 
\label{FSinB}
\end{center}
\end{figure}

The dimensional reduction of the LLL which is so important for the chiral condensation in the mechanism of MC is irrelevant at finite density, as the excitations about the Fermi surface are already 1+1-dimensional ((1+1)-D) at zero field, because their energy only changes in the direction perpendicular to the Fermi surface. However, a magnetic field can affect the pairing mechanism in a different way. In the presence of a magnetic field, the geometry of the Fermi surface changes, turning into a discrete set of rings defined by the intersection of the surface of the Fermi sphere at zero field with the cylinders associated with the different Landau levels (Fig.\ref{FSinB}) in momentum space. Pairing now can occur between particles excited in small cylinders about each Landau level in the Fermi surface. Therefore, the field influences the pairing mechanism by this modification of the Fermi surface, and also through the change in the degeneracy of the states, $2\int_{-\infty}^{\infty}\frac{d^{3}p}{(2\pi)^{3}}\rightarrow\frac{|qB|}{2\pi}{\displaystyle \sum_{l=0}^{\infty}}\left(2-\delta_{l0}\right)\int_{-\infty}^{\infty}\frac{dp_{3}}{2\pi}$, which now becomes proportional to the field. 

 As I shall discuss in the next sections,  in the context of QCD a magnetic field can affect the realization of Cooper and DW pairings in quite nontrivial ways. 

\section{Magnetic CFL Superconductivity}
Cooper pairing begins to be important for QCD in the high density, low temperature region, where it is responsible for the phenomenon of color superconductivity. Because pairing of two quarks is always colored, the ground state breaks the color symmetry forming a color superconductor. The most favored phase at very high densities is the CFL phase that is realized through the color-antitriplet, flavor-antitriplet interaction channel. The CFL ground state breaks chiral symmetry through a locking of color and flavor transformations and reduces the original symmetry to the diagonal subgroup $SU(3)_{C+L+R}$ of locked transformations \cite{ASRreview}. This unbroken group contains an Abelian $U(1)_{\widetilde{Q}}$ subgroup which consists of a simultaneous electromagnetic and color rotation and plays the role of a "rotated" electromagnetism. The group generator $\widetilde{Q}$  remains unbroken because all the diquarks in the condensate have zero $\widetilde{Q}$-charge. Hence, the $\widetilde{Q}$ photon is massless and consequently, a rotated magnetic field will not be subject to Meissner effect in the CFL superconductor. Since the mixing angle between the original electromagnetic and gluon generators is very small, the $\widetilde{Q}$ photon is mostly the original photon with a small admixture of gluon. Thus, a regular magnetic field will penetrate a CFL superconductor almost unabated.

\begin{figure}[ht!]
\begin{center}
\includegraphics[width=0.5\textwidth]{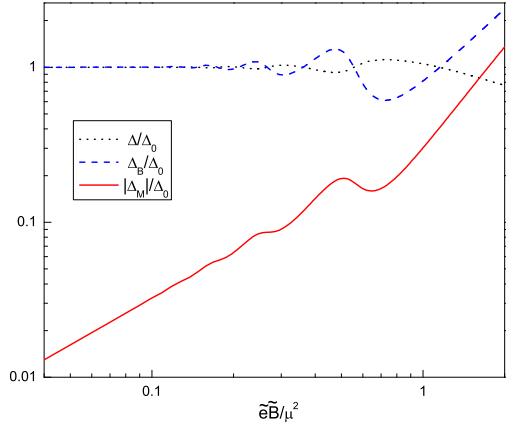}
\caption{\footnotesize Evolution of the MCFL gaps with the magnetic field}
\label{MCFLgaps}
\end{center}
\end{figure}

Although all the diquarks have zero net $\widetilde{Q}$-charge, some are formed by neutral constituents (both quarks $\widetilde{Q}$-neutral) and some by charged constituents (quarks in the pair have opposite $\widetilde{Q}$-charge). The $\widetilde{Q}$-charged quarks in the last set of pairs can couple to an external magnetic field and lead to a splitting of the CFL gap $\Delta_{CFL}$ into a gap $\Delta$ that only gets contribution of diquarks with neutral quarks, and a second gap $\Delta_B$ that gets contributions of diquarks with $\widetilde{Q}$-neutral and $\widetilde{Q}$-charged constituents \cite{MCFL}. In addition, a third gap $\Delta_M$ is also formed because the Cooper pairs with $\widetilde{Q}$-charged constituents have nonzero AMM \cite{MCFL-AMM}. Similarly to what occurs in the MC scenario, the explicit breaking of the rotational symmetry by the uniform magnetic field opens new channels of interactions through the Fierz transformations and allows the formation of a spin-one diquark condensate characterized by the gap $\Delta_M$. This new order parameter is proportional to the component in the field direction of the average magnetic moment of
the pairs of charged quarks. As can be seen in Fig.\ref{MCFLgaps}, in the region of large fields, the magnitude of  $\Delta_M$ is bigger than $\Delta$ and comparable to $\Delta_B$. Since there is no solution of the gap equations with nonzero scalar gaps and zero value of this magnetic moment condensate, its presence in the MCFL phase is unavoidable.  At lower fields, the MCFL phase exhibits de typical Haas-van Alphen oscillations \cite{OscilationMCFL} of magnetized systems.

Even though the separation between the $\Delta$ and $\Delta_B$ gaps is relevant only at very strong fields,  the difference between the low-energy physics of the CFL and the MCFL phases becomes important at much smaller field strengths, of order $\Delta^2_{CFL}$ \cite{BPhases}. This can be understood from the following considerations:  

In the CFL phase the symmetry breaking is 
\begin{equation}\label{CFL}
SU(3)_C \times SU(3)_L \times SU(3)_R \times U(1)_B
\rightarrow SU(3)_{C+L+R}\times
Z_2.
\end{equation}
This symmetry reduction leaves nine Goldstone bosons: a singlet
associated to the breaking of the baryonic symmetry $U(1)_B$, and
an octet associated to the broken axial group $SU(3)_A$.

Once a magnetic field is switched on, the difference between the
electric charge of the $u$ quark and that of the $d$ and $s$
quarks reduces the original flavor symmetry of the theory and also the symmetry group that remains after the diquark
condensation. Then, the breaking pattern for the
MCFL-phase \cite{MCFL} becomes
\begin{equation}\label{MCFL}
SU(3)_C \times SU(2)_L \times SU(2)_R \times
U(1)^{(1)}_A\times U(1)_B  \rightarrow
SU(2)_{C+L+R} \times Z_2.
\end{equation}
The group $U(1)^{(1)}_A$ (not to be confused with the usual
anomaly $U(1)_{A}$) is related to the current which is an
anomaly-free linear combination of $s$, $d$, and $u$ axial
currents. In this case only five
Goldstone bosons remain. Three of them correspond to the breaking
of $SU(2)_A$, one to the breaking of $U(1)^{(1)}_A$, and one to
the breaking of $U(1)_B$. Thus, an applied magnetic field reduces
the number of Goldstone bosons in the superconducting phase, from
nine to five.

Not only has the MCFL phase a smaller number of
Goldstone fields, but all these bosons are
neutral with respect to the rotated electric charge. Hence, no
charged low-energy excitation can be produced in the MCFL phase.
This effect can be relevant for the low energy physics of a color
superconducting star's core and hence for the star's transport
properties. In particular, the cooling of a compact star is
determined by the particles with the lowest energy; so a star with
a core of quark matter and sufficiently large magnetic field could display a distinctive cooling process.

Although the symmetries of the CFL and MCFL ground states are quite different, at weak fields the CFL phase still remains as a good approximation to describe the low energy physics, since the masses of the charged Goldstone bosons depend on the magnetic field and are very small. However, as shown in Ref. \cite{BPhases}, when the field increases and becomes comparable to $\Delta^2_{CFL}$, the mass of the charged Goldstone bosons is large enough for them to decay into a particle-antiparticle pair, and only the neutral Goldstone bosons remain. This is the energy scale at which the MCFL phase becomes physically relevant.

The MCFL matter is subject to magnetoelectricity, pressure anisotropies, and several other interesting effects. For a review see \cite{MCFLreview}. The possibility of self-bound MCFL matter in neutron stars was explored in \cite{MCFLEoS}, where the magnetic field was found to act as a destabilizing factor for the realization of strange matter in such a way that only if the bag constant decreases with the field, a magnetized strange star could exist.

\section{Quarkyonic Matter in a Magnetic Field}
In this section, we will consider fermion pairing in QCD at low temperatures and intermediate densities in the large $N_c$ limit. In the region of intermediate densities, i.e., large enough for the system to be in the quark phase, but small enough to support nonperturbative interactions, color superconductivity 
and DW pairing compete with each other. In the large $N_c$ limit the diquark condensate is definitely not favored because it is not a color singlet and decreases as $1/N_c$. These are the conditions where quarkyonic matter can be realized. 

Quarkyonic matter (QyM) is a large $N_c$ phase of cold dense quark matter recently suggested in Ref. \cite{quarkyonic-matter}. The main feature of QyM is the existence of asymptotically free quarks deep in the Fermi sea and confined excitations at the Fermi surface. The quarks lying deep in the Fermi sea are weakly interacting because they are hard to be excited due to Pauli blocking. Their interactions are hence very energetic and the confining part of the interaction does not play any role \cite{1105.4103}. On the other hand, excitations of quarks within a shell of width $\Lambda_{QCD}$ from the Fermi surface interact through infrared singular gluons at large $N_c$ and hence are confined. For the QyM to exist, the screening effects have to be under control, so they cannot eliminate confining at the Fermi surface. Such a region can be defined by the condition $m_{D}\ll \Lambda_{QCD}\ll\mu$, with $m_{D}$ the screening mass of the gluons and $\mu$ the quark chemical potential \cite{chiralspirals-conf}.

As shown in Refs. \cite{chiralspirals-conf,chiralspirals}, chiral symmetry can be broken in QyM through the formation of a translational non-invariant condensate that arises from the pairing between a quark with momentum $\mathbf{P}$ and the hole formed by removing a quark with opposite momentum $\mathbf{-P}$ from the Fermi surface. The DW condensate that forms in QyM is a linear combination of the chiral condensate $\langle\overline{\psi}\psi\rangle$, and a spin-one, isosinglet odd-parity condensate of $\langle\overline{\psi}\sigma^{0z}\psi\rangle$. Here z is the direction of motion of the wave. At each given patch of the Fermi surface, z is the direction perpendicular to that surface. This combination of two inhomogeneous condensates has been named  Quarkyonic Chiral Spiral (QyCS) \cite{chiralspirals}. The $\langle\overline{\psi}\sigma^{0z}\psi\rangle$ component corresponds to the condensation of an electric dipole moment. 

\begin{figure}[ht!]
\begin{center}
\includegraphics[width=0.4\textwidth]{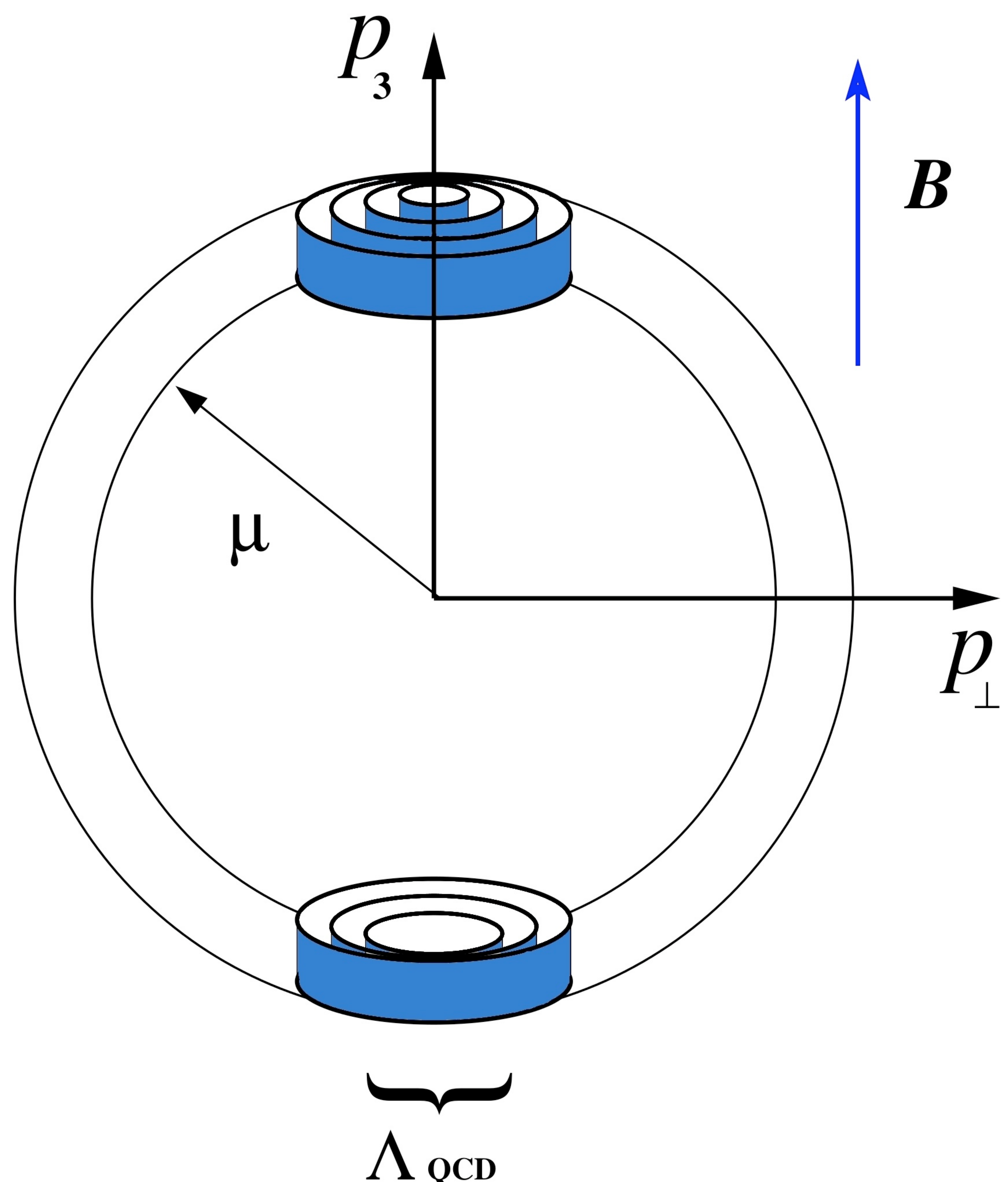}
\caption{\footnotesize Polar patches in the Fermi surface at nonzero magnetic field.}
\label{FSPatches}
\end{center}
\end{figure}

The influence of a magnetic field on the QyCS was investigated in Ref. \cite{QySB} using a single-flavored (3+1)-D QCD theory with a Gribov-Zwanziger confined gluon propagator. Considering the polar patches shown in Fig. \ref{FSPatches}, and assuming a magnetic field that points in the z-direction and has magnitude $B\le\Lambda^{2}_{QCD}$, it was shown found that the (3+1)-D theory is mapped into the following (1+1)-D QCD theory
\bea
   \emph{L}^{2D}_{eff}&=&
     \overline{\Phi}_0
     [i\Gamma^\mu(\partial_\mu+i g_{2D}A_\mu +\Gamma^0 \mu]\Phi_0
     \nonumber \\
    &+&\sum_{l=1}^{L}
     \overline{\Phi}_l[i\Gamma^\mu(\partial_\mu+ ig_{2D}A_\mu)+\Gamma^0\mu]\Phi_l -\frac{1}{2}trG^{2}_{\mu\nu},
\label{2DlagwithB}
\eea
with $2L+1$ flavors and flavor symmetry SU(2L)$\times$ U(1).

The spinor fields in this (1+1)-D theory are defined by $\Phi^{T}_0=( \varphi^{(0)}_{\uparrow},0)$ and $\Phi^{T}_l=( \varphi^{(l)}_{\uparrow},\varphi^{(l)}_{\downarrow})$, with flavor indexes $l$, and $\uparrow$, $\downarrow$  corresponding respectively to the Landau level and the spin up and down components of the 4D spinors of the original 4D theory. The 2D Dirac $\Gamma$ matrices are defined in term of the Pauli matrices as $\Gamma^0=\sigma^1$; $\Gamma^z=-i\sigma^2$; $\Gamma^5=\sigma^3$. $L$ is the maximum number of Landau levels that can fit into the polar patches.

Performing the following transformation of the quark fields
\bea
\Phi_{l}=exp(-i\mu z \Gamma_{5})\Phi'_{l} \quad\quad l=0,...L
\label{fieldtransf}
\eea
to eliminate the chemical potential (it actually remains in the theory through the anomaly of the baryon charge) one obtains
\bea
   \emph{L}^{2D}_{eff}&=\sum_{l=0}^{L}
     \overline{\Phi}'_{l}[i\Gamma^\mu(\partial_\mu+ ig_{2D}A_\mu)]\Phi'_{l} -\frac{1}{2}trG^{2}_{\mu\nu}
\label{transf2DlagwithB}
\eea

As argued in \cite{QySB}, the above theory admits the formation of  two independent condensates,
\bea
 \langle \overline{\Phi}'\Phi' \rangle 
 = \langle \overline{\varphi}^{'(0)}_{\uparrow}\varphi^{'(0)}_{\uparrow} \rangle +\sum_{l=1}^{L}[ \langle \overline{\varphi}^{'(l)}_{\uparrow}\varphi^{'(l)}_{\uparrow} \rangle+\langle \overline{\varphi}^{'(l)}_{\downarrow}\varphi^{'(l)}_{\downarrow} \rangle]
\label{Ch-cond-decomp1}
\eea
and 
\bea
 \langle \overline{\Phi}'\tau_3 \Phi' \rangle
 = \langle \overline{\varphi}^{'(0)}_{\uparrow}\varphi^{'(0)}_{\uparrow} \rangle +\sum_{l=1}^{L}[ \langle \overline{\varphi}^{'(l)}_{\uparrow}\varphi^{'(l)}_{\uparrow} \rangle-\langle \overline{\varphi}^{'(l)}_{\downarrow}\varphi^{'(l)}_{\downarrow} \rangle]
\label{Ch-cond-decomp2}
\eea
See Ref. \cite{chiralspirals} for the definition of the flavor matrix $\tau_3$ in the present context. A condensate of the form $\langle \overline{\Phi}'\tau_3 \Phi' \rangle=\langle \overline{\varphi}'_{\uparrow}\varphi'_{\uparrow} \rangle-\langle \overline{\varphi}'_{\downarrow}\varphi'_{\downarrow} \rangle$ is not present in the QyM at zero magnetic field because at zero field there is a spin degeneracy so spin up and down condensates have to be the same and thus cancel out in $\langle \overline{\Phi}'\tau_3 \Phi' \rangle$, in agreement with the claims of Ref. \cite{chiralspirals}. When $B \neq0$, the LLL contribution, which is the only level that has no spin degeneracy, makes it possible for this second condensate to be present.This is true even if the two spin-flavor terms in the sum (\ref{Ch-cond-decomp2}) cancel out. 

In terms of the unprimed fields we find
\bea
 \langle \overline{\Phi}\Phi \rangle
 =\cos(2\mu z)\langle \overline{\Phi}'\Phi' \rangle, \quad \langle \overline{\Phi}\Gamma_5 \Phi \rangle
 =-i\sin(2\mu z)\langle \overline{\Phi}'\Phi' \rangle,
\label{spirals1}
\eea
and 
\bea
 \langle \overline{\Phi}\tau_3\Phi \rangle
 =\cos(2\mu z)\langle \overline{\Phi}'\tau_3\Phi' \rangle, \quad \langle \overline{\Phi}\tau_3\Gamma_5 \Phi \rangle
 =-i\sin(2\mu z)\langle \overline{\Phi}'\tau_3 \Phi' \rangle.
\label{spirals2}
\eea
Therefore, in the presence of a magnetic field two set of inhomogeneous condensates emerge. 

Here again an extra condensate is generated thanks to the explicit breaking of the rotational symmetry by the magnetic field. Notice that the matrix $\tau_3$ is a generator of the group of flavor symmetries in (1+1)D, but in the 4D theory, it is actually related to the spin component in the third direction, which is precisely the direction of the external magnetic field.

Going back to the quark fields in the 4D theory, the two chiral spirals in the presence of a magnetic field are
\bea
 \langle \overline{\psi}\psi \rangle
 =\Delta_1\cos(2\mu z), \quad \quad \langle \overline{\psi}\gamma^0\gamma^3\psi \rangle
 =\Delta_1\sin(2\mu z)
\label{4Dspirals1}
\eea
\bea
 \langle \overline{\psi}\gamma^1\gamma^2\psi \rangle
 =\Delta_2 \cos(2\mu z), \quad \quad \langle \overline{\psi}\gamma^5\psi \rangle
 =\Delta_2 \sin(2\mu z)
\label{4Dspirals2}
\eea

The field-induced chiral spiral is a combination of a condensate of magnetic moment and a pion condensate, also varying in the direction parallel to the field. Both parity and time-reversal symmetries are broken in the system. The spontaneous generation of inhomogeneous condensates with electric and magnetic dipole moments may lead to interesting observational implications with potential consequences in dense environments like the cores of neutron stars or the planed high-density heavy-ion collision experiments. 

\bigskip
I am very grateful to the local organized committee of CSQCDIII for the invitation and warm hospitality. This work has been supported in part by DOE Nuclear Theory grant DE-SC0002179.

\end{document}

%% file: econfmacros-csqcd3.tex
\def\beq{\begin{equation}}
\def\eeq#1{\label{#1}\end{equation}}
\def\eeqn{\end{equation}}

\def\beqa{\begin{eqnarray}}
\def\eeqa#1{\label{#1}\end{eqnarray}}
\def\eeqan{\end{eqnarray}}

\let\bar=\overbar

\def\Dslash{\not{\hbox{\kern-4pt $D$}}}
\def\dslash{\not{\hbox{\kern-2pt $\del$}}}

\def\msb{{\bar{\ssstyle M \kern -1pt S}}}

\usepackage{fancyhdr,graphicx}